\RequirePackage{ifpdf}
\documentclass[hyper]{JHEP3}
\pdfoutput=1
\usepackage{graphicx}
\usepackage{epsfig}
\usepackage{multicol}
\usepackage{bbm}
\usepackage{amsmath}
\usepackage{cite}
\usepackage{lscape}
\usepackage{multirow}
\usepackage{amscd} 
\usepackage{slashed}

\newcommand{\smodels}{{\texttt{SModelS}}}
\newcommand{\axe}{$(A\times\epsilon)$ }
\newcommand{\eaxe}{(A\times\epsilon) }

\allowdisplaybreaks

\title{Simplified models for same-spin new physics scenarios}

\author{
Lisa Edelh\"auser${}^a$, Michael Kr\"amer${}^{a,b}$ and Jory Sonneveld${}^a$\\
${}^a$Institute for Theoretical Particle Physics and Cosmology, RWTH Aachen University,\\ \hspace*{1.75mm}52056 Aachen, Germany

\vspace*{1mm}

${}^b$SLAC National Accelerator Laboratory, Stanford University, Stanford, CA 94025, USA}

\abstract{
Simplified models are an important tool for the interpretation of searches for new physics at the LHC. They are 
defined by a small number of new particles together with a specific production and decay pattern. The simplified 
models adopted in the experimental analyses thus far have been derived from supersymmetric theories, 
and they have been used to set limits on supersymmetric particle masses. We investigate the 
applicability of such simplified supersymmetric models to a wider class of new physics scenarios, in particular 
those with same-spin Standard Model partners. We focus on the pair production of quark partners
and analyze searches for jets and missing energy within a simplified supersymmetric 
model with scalar quarks and a simplified model with spin-1/2 quark partners. Despite sizable differences in the 
detection efficiencies due to the spin of the new particles, the limits on particle masses are 
found to be rather similar. We conclude that the supersymmetric simplified models employed in current experimental analyses 
also provide a reliable tool to constrain same-spin BSM scenarios.}


\preprint{
SLAC-PUB-16176\\
TTK-15-01} 

\begin{document}
\section{Introduction}

Searches for new physics at the LHC are often interpreted in terms of simplified models \cite{Alves:2011wf,Alwall:2008ag,Chatrchyan:2013sza}. 
Simplified models are characterised by a small number of parameters and a simple production and decay pattern. It is 
assumed that the model-specific details of the dynamics of production and decay have little influence on the signal efficiencies, so that 
results from simplified model analyses can be applied to a wide range of new physics scenarios. 
Recently developed program packages provide a convenient framework to employ simplified models for testing BSM theories \cite{Kraml:2013mwa,Kraml:2014sna,Papucci:2014rja} at the LHC\@.

The simplified models that have been used in recent ATLAS and CMS new physics searches are derived from supersymmetric 
models. Thus, they include scalar quark partners and fermionic partners of the Standard Model (SM) gauge bosons. Non-supersymmetric extensions of the SM, like 
little Higgs models or models with universal extra dimensions, on the other hand, include same-spin SM partners. The spin of the new particles affects 
their kinematic distributions and thus the detection efficiencies. Therefore, it is important to investigate the impact of the particle spin on the 
exclusion limits, and to quantify to which extent the current analyses based on supersymmetric simplified models can be used to constrain beyond the Standard Model (BSM) scenarios with same-spin partners. 

An important signature for new physics searches at the LHC is dijet production with missing transverse energy (MET). Such a signature can be described 
by the simplified model T2\cite{Chatrchyan:2013sza}, which corresponds to supersymmetric squark-antisquark production in the limit of a heavy gluino. T2 includes up and down 
squarks that directly decay into a light-flavor jet and the lightest neutralino. We compare the efficiencies and mass limits as derived from the supersymmetric 
simplified model T2 with efficiencies and mass limits derived from a model with same-spin quark partners. Specifically, we define a simplified model with spin-1/2 quark partners based on universal extra dimensions, as detailed in section~\ref{sec:t2setup}, and reinterpret experimental searches for jets and MET in such a model. 

To identify the most sensitive experimental analyses for constraining quark-partner pair production, we employ the public tool \smodels\cite{Kraml:2013mwa, Kraml:2014sna} (which uses \cite{Buckley:2013jua, Beenakker:1996ch, Sjostrand:2006za}). 
\smodels\ decomposes a generic BSM collider signature into simplified model topologies, and confronts these topologies with the relevant experimental constraints from ATLAS and CMS. We find that the CMS $\alpha_T$ \cite{Chatrchyan:2013lya}, MHT \cite{Chatrchyan:2014lfa}, and effective mass $M_{T2}$ \cite{CMS-PAS-SUS-13-019} analyses, as well as one ATLAS analysis \cite{ATLAS-CONF-2013-047}, provide exclusions for UED-like quark production. In this work, we focus on the CMS analyses \cite{Chatrchyan:2013lya} and \cite{Chatrchyan:2014lfa} as representative examples. 

This paper is structured as follows. In section \ref{sec:t2setup} we define the simplified models based on supersymmetry and universal extra dimensions, respectively. 
In section \ref{sec:searches} we introduce the two CMS searches. The comparison of efficiencies and quark-partner mass limits is presented in section~\ref{sec:results}. We conclude in section \ref{sec:conclusions}.

\section{Supersymmetric and same-spin simplified models}\label{sec:t2setup}
\begin{figure}[htp]\centering
\includegraphics[width=0.15\textwidth]{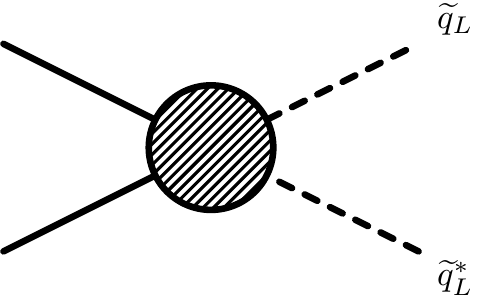}\hspace{2ex}\raisebox{3ex}{=\hspace{2ex}$\Bigg\{$}
\includegraphics[width=0.15\textwidth]{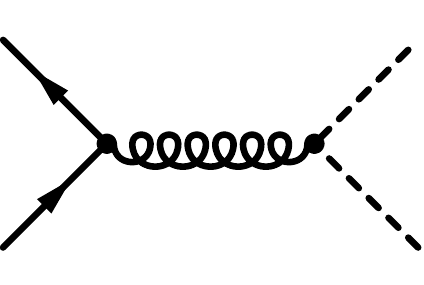}
\includegraphics[width=0.15\textwidth]{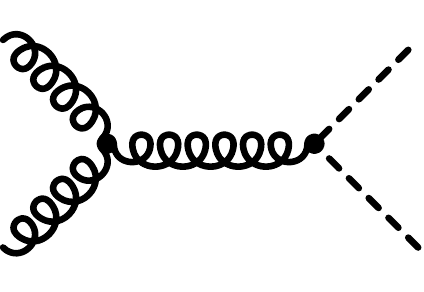}
\includegraphics[width=0.15\textwidth]{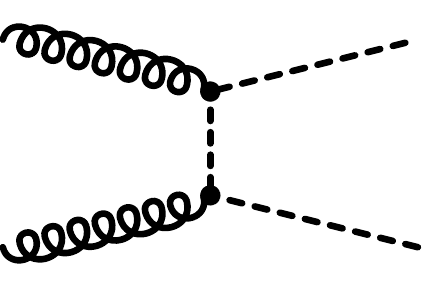}
\includegraphics[width=0.15\textwidth]{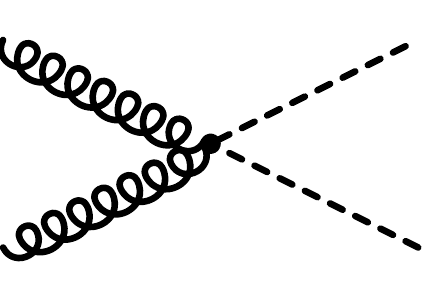}\raisebox{3ex}{$\Bigg\}$}
\caption{The diagrams contributing to squark pair production in the T2 supersymmetric simplified model.}\label{fig:T2susy}
\end{figure}

The simplified model T2 used by  the ATLAS and CMS collaborations to interpret searches for jets and MET corresponds to the production of 
a squark-antisquark pair 
\begin{align}
  pp \rightarrow \tilde{q}_i \bar{\tilde{q}}_i, ~~~~~~~\mbox{with chirality  }i\in\{L,R\}
  \label{eq:t2def}
\end{align}
with a decoupled gluino $m_{\tilde{g}} \to \infty$. Each squark directly decays to a quark and a neutralino, where the latter is assumed to be the lightest supersymmetric particle (LSP): $\tilde{q} \to q \,\tilde{\chi}_{\rm LSP}$. 
In realistic supersymmetric models, the gluino is in general not decoupled and additional production channels mediated by gluino exchange contribute to the squark cross section, such as $pp \rightarrow \tilde{q}_i \tilde{q}_i$, $pp \rightarrow \tilde{q}_i \bar{\tilde{q}}_j$ and $pp \rightarrow \tilde{q}_i \tilde{q}_j$ with $i\neq j$. These can compete with and exceed the production mode of T2 (\ref{eq:t2def}), see e.g.\ the discussion in Ref.\cite{Edelhauser:2014ena}. In figure \ref{fig:T2susy}, the diagrams contributing to squark pair production in the supersymmetric T2 model are shown. 

As a concrete example of a model with same-spin SM partners we consider universal extra dimensions (UED) \cite{Appelquist:2000nn,Hooper:2007qk}. UED models are the simplest extension of the Standard Model including one compactified extra dimension. All fields are assumed to propagate in a flat 4+1 dimensional space-time. Due to the compactification of the extra dimension, the momentum along the 5th dimension is discretized, leading to a tower of Kaluza-Klein (KK) resonances. Each SM particle is therefore accompanied by a tower of KK excitations with the same quantum numbers, in particular the same spin, but with larger masses. 
An additional feature of the model, KK-parity, is a $Z_2$ symmetry in the extra dimension. It guarantees the stability of the lightest KK-excitation (LKP) providing a viable dark matter candidate  \cite{Servant:2002aq,Servant:2002hb}.

As a result of KK-parity, when considering only the first (and lightest) KK mode, the collider phenomenology of UED models can be qualitatively similar to that of the minimal supersymmetric Standard Model \cite{Nilles:1983ge}. This allows us, in principle, to interpret searches for supersymmetry in jets and MET final states within UED models and to set limits on UED particle masses, see also \cite{Cacciapaglia:2013wha, Datta:2011vg}. 
However, as mentioned in the introduction, because of the different spins of the supersymmetric and UED particles, the signal efficiencies are different in general, and it is thus not clear a priori to what extent limits derived for supersymmetric models can be applied to UED.

While the masses of the KK excitations in UED models are, in general, determined by the compactification radius of the extra dimension, we take  them to be free parameters, in accordance with the simplified model approach. Our model should thus be viewed as a same-spin toy model, designed to explore the differences with supersymmetry due to spin effects. To calculate the cross sections and distributions we use \texttt{MadGraph 5.1.12} \cite{Alwall:2014hca} together with our own extension \cite{Edelhauser:2013lia} of the implementation of the minimal UED model in \texttt{Feynrules} \cite{Christensen:2008py,Christensen:2009jx}. For more details on the UED model see 
e.g.\ Ref.\cite{Datta:2010us}.

\begin{figure}[htp]\centering
\includegraphics[width=0.15\textwidth]{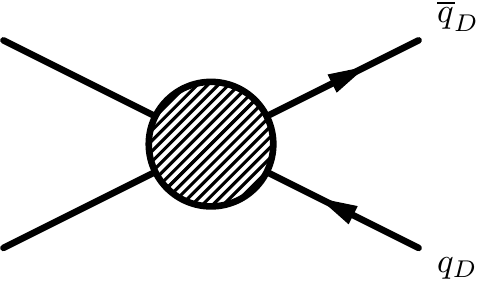}\hspace{2ex}\raisebox{3ex}{=\hspace{2ex}$\Bigg\{$}
\includegraphics[width=0.15\textwidth]{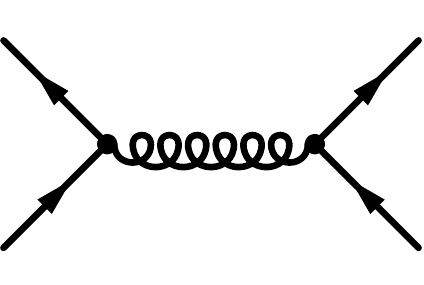}
\includegraphics[width=0.15\textwidth]{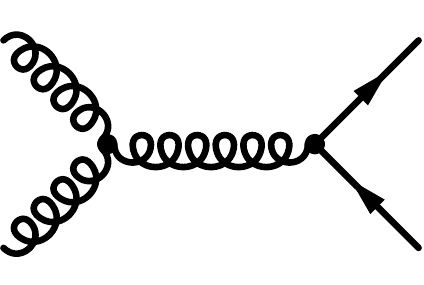}
\includegraphics[width=0.15\textwidth]{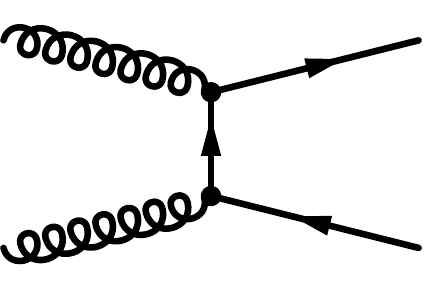}\raisebox{3ex}{$\Bigg\}$}
\caption{The diagrams contributing to KK quark pair production in the UED-T2 simplified model.}\label{fig:T2ued}
\end{figure}

The quark partners in our UED model are the first KK excitations of the quark doublets and singlets, denoted by $q_D^{(1)}$ and $q_S^{(1)}$, respectively. The dark matter candidate $B^{(1)}$ is the first KK excitation of the Standard Model U(1) gauge field.
To define the UED-T2 simplified model, we consider the pair production of KK quark doublets and singlets  
\begin{align}
pp \rightarrow q_i^{(1)} \overline{q}_i^{(1)}
\end{align}
where $i\in\{D,S\}$, with the KK gluon decoupled. The KK quarks decay directly to quarks and the LKP: $q_i^{(1)}\rightarrow q\,B^{(1)}$. The corresponding diagrams for  KK quark pair production in the UED-T2 model are shown in figure \ref{fig:T2ued}.

In more realistic UED models, the KK gluon is not decoupled from the KK particle spectrum but has a finite mass, and additional production channels contribute to the KK quark cross section:
\begin{align}\label{eq:ued_prod}
pp \rightarrow q_i^{(1)} q_i^{(1)},~~~
pp \rightarrow q_i^{(1)} q_j^{(1)},~~~
pp \rightarrow q_i^{(1)} \overline{q}_j^{(1)}\,,
\end{align}
where $i, j\in\{D,S\}, i\neq j$. To simplify notation, we drop the superscript  $^{(1)}$ for the first KK mode in most of the following.

\begin{figure}[]
  \begin{center}
    \includegraphics[width=0.49\textwidth]{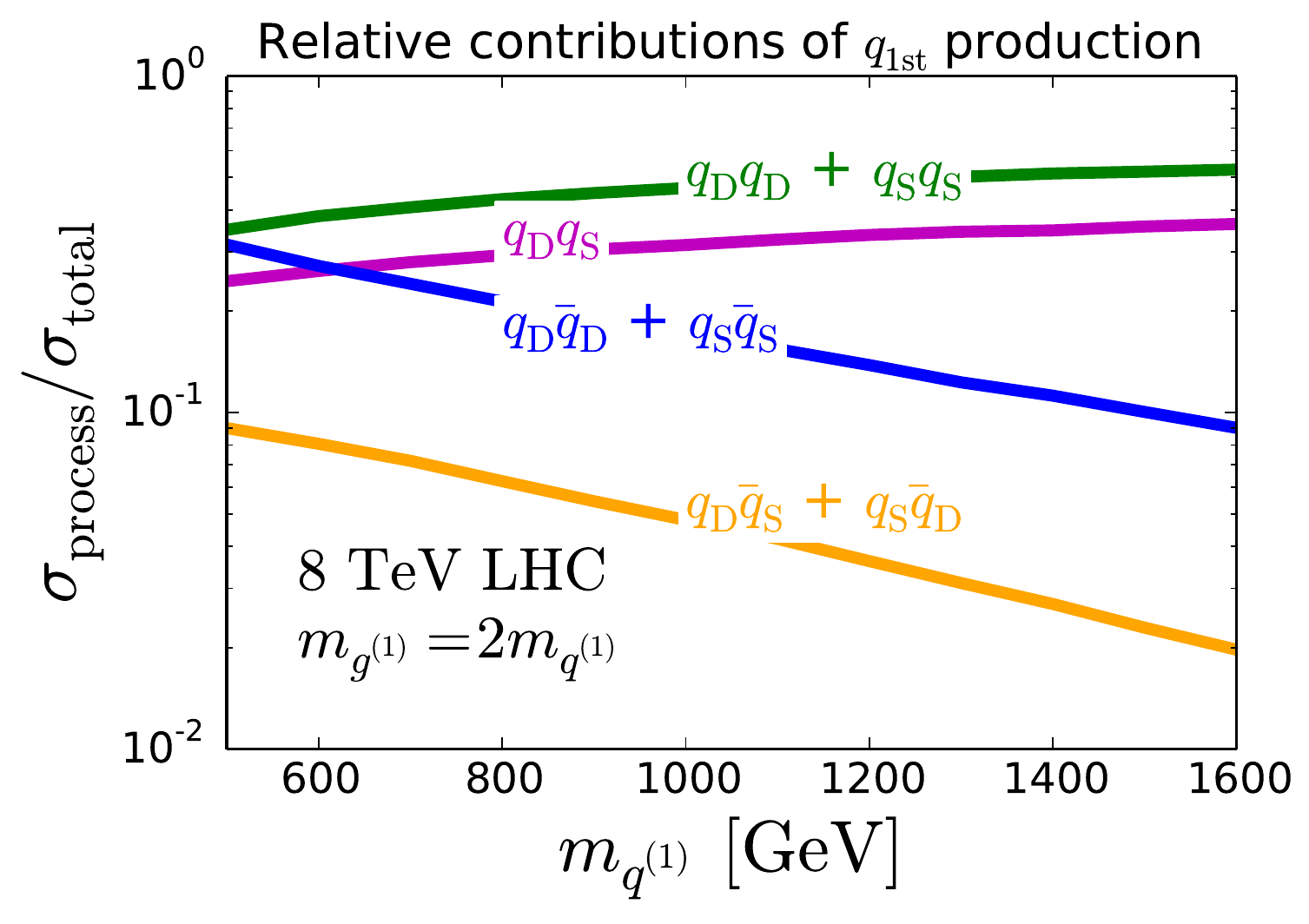}
    \includegraphics[width=0.49\textwidth]{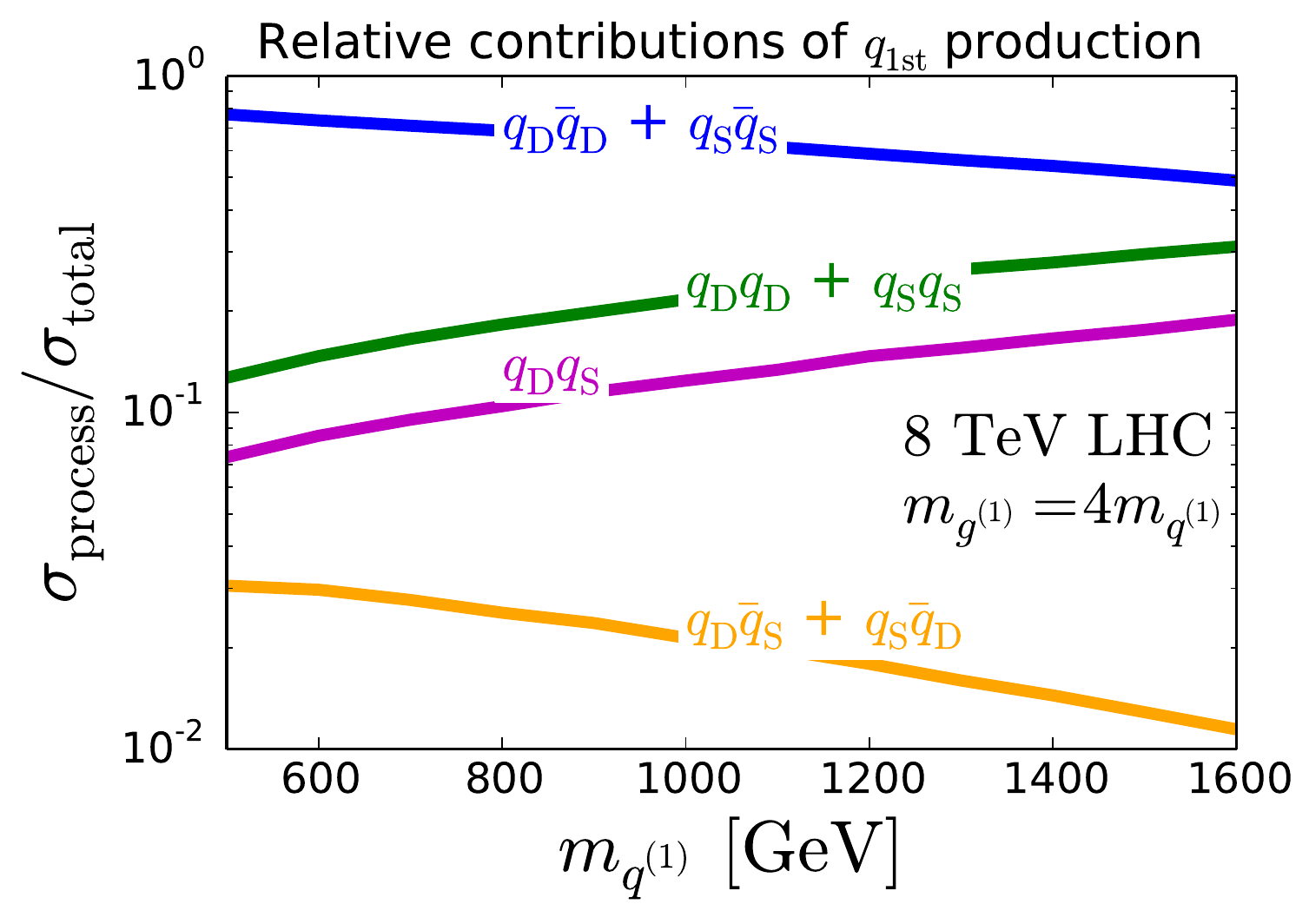}
  \end{center}
  \vspace*{-5mm}
  \caption{Relative production cross sections for the first KK excitation singlet and double quarks in a minimal universal extra dimension model. Only KK quarks of the first generation are taken into account. The KK gluon mass is set to twice (left plot) and four times (right plot) the KK quark mass. Cross sections of equally contributing production processes were added.}
  \label{fig:kkquarkproduction1st}
\end{figure}

In figure \ref{fig:kkquarkproduction1st} we show the relative contributions of the various KK quark production channels as a function of the KK quark mass for two different choices of the KK gluon mass. 
\begin{figure}[]
  \begin{center}
    \includegraphics[width=0.49\textwidth]{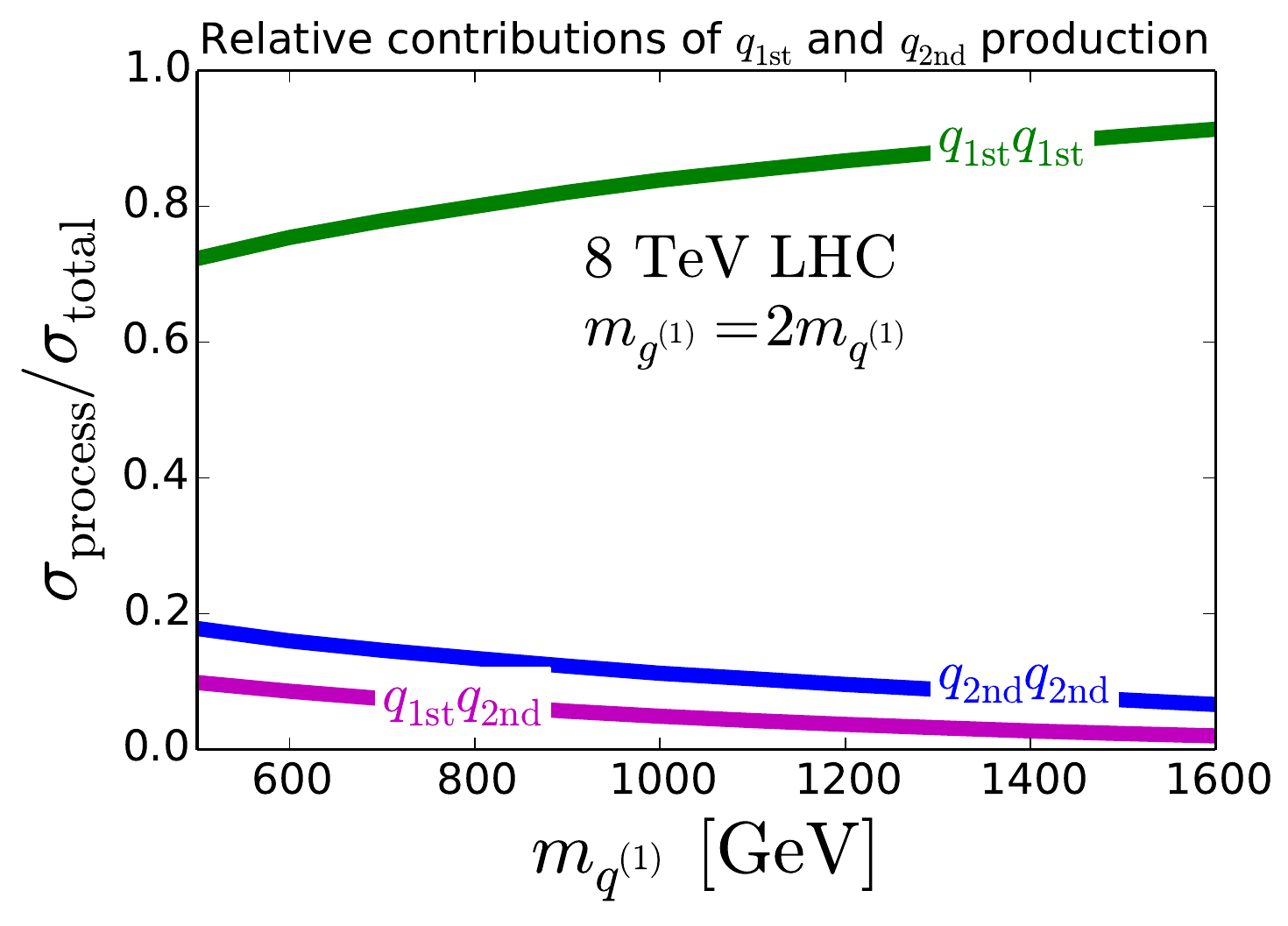}
    \includegraphics[width=0.49\textwidth]{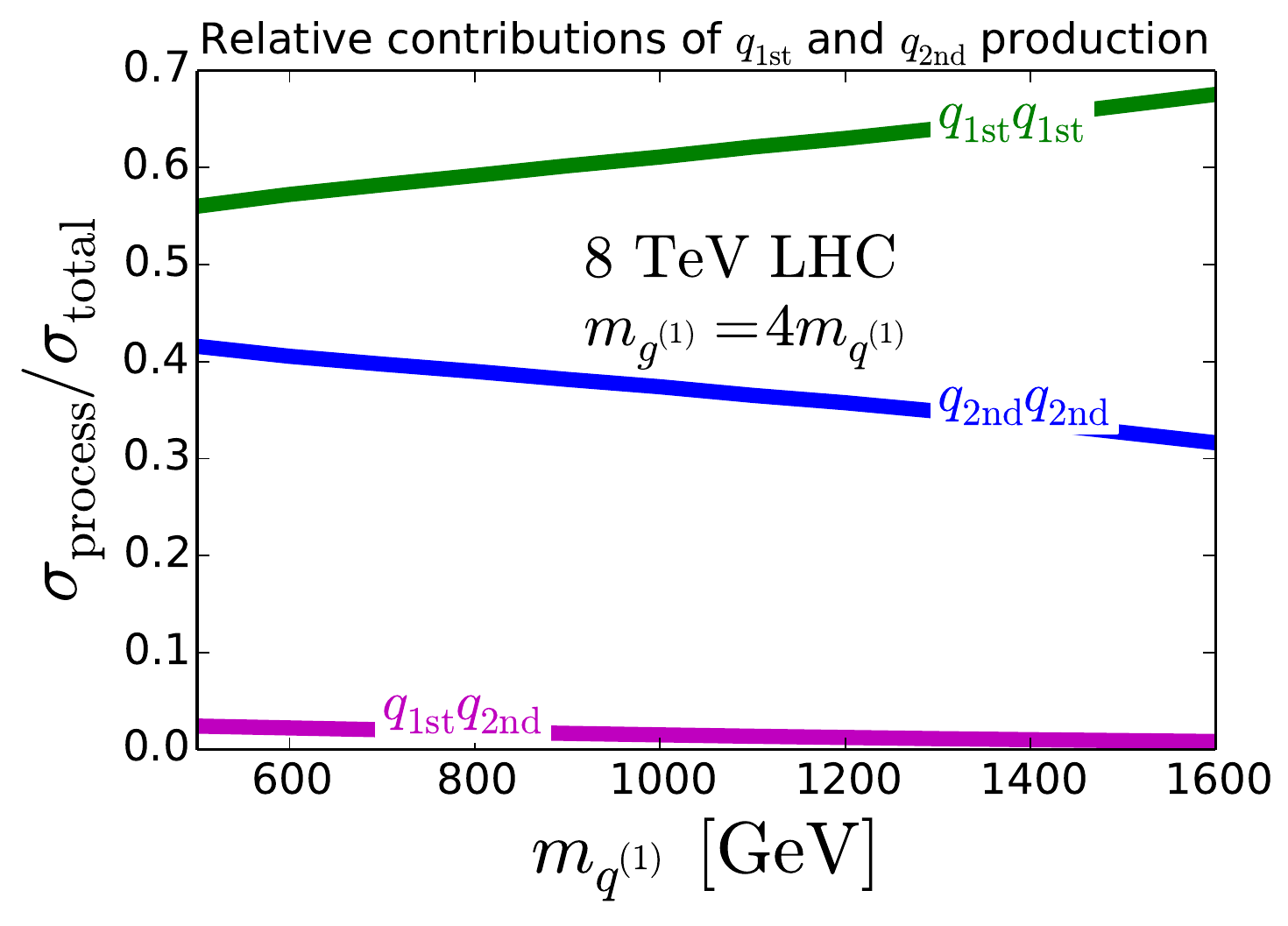}
  \end{center}
    \vspace*{-5mm}
  \caption{Relative production cross sections for the first KK quark excitations of the first and second generation. The cross sections include contributions from both singlet and doublet KK quarks.  The KK gluon mass is set to twice (left plot) and four times (right plot) the KK quark mass.}  
  \label{fig:kkquarkproduction1st2nd}
\end{figure}
For a relatively light KK gluon, the production processes $q_D \overline{q}_D$ and $q_S \overline{q}_S$ are subdominant to those of $q_D q_S$ and $q_D q_D$. For a heavier KK gluon the situation is reversed, and the KK quark-antiquark production processes $q_D \overline{q}_D$ and $q_S \overline{q}_S$ are dominant.

The relative importance of first and second generation KK quark production (summing singlet and doublet KK quarks) is shown in figure \ref{fig:kkquarkproduction1st2nd} 
for two KK-gluon masses, $m_{g^{(1)}}/m_{q^{(1)}}\in \{2,4\}$. In the case of a relatively light KK gluon, the production of two first generation KK quarks is dominant (left plot), while in the case of a heavier KK gluon, the production of two second generation KK quarks becomes important as well.

\section{Searches and setup}\label{sec:searches}
In this section we briefly describe two experimental searches that are sensitive to the simplified model T2 and that were used in this study 
to test the accuracy of the SUSY-T2 model for limits on same-spin BSM models:  the analysis of Ref.\cite{Chatrchyan:2013lya}, which is based on the variable $\alpha_T$, and the analysis of Ref.\cite{Chatrchyan:2014lfa}, a search based on missing transverse momentum (MHT). We also describe the setup and tools used for our simulation to reinterpret the experimental searches. 

\paragraph{The $\alpha_T$ search \cite{Chatrchyan:2013lya}}    The variable $\alpha_T$ is for dijet events defined as 
    \begin{equation}
      \alpha_T = \frac{E_T^{j_2}}{M_T},
      \label{eqn:alphaT_dijet}
    \end{equation}
    where $E_T^{j_2}$ is the transverse energy of the second hardest jet, and $M_T$ is the transverse mass:
     \begin{equation}
      M_T = \sqrt{\left(\sum^{N_{\mathrm{jet}}}_{i =1} E_T^{j_i}\right)^2 - \left(\sum^{N_{\mathrm{jet}}}_{i =1} p_x^{j_i}\right)^2 - \left(\sum^{N_{\mathrm{jet}}}_{i =1} p_y^{j_i}\right)^2}.
      \label{eqn:MTransverse}
    \end{equation}
    For events with more than two hard jets, all jets are combined into two pseudojets such that the difference between the scalar sum of the transverse energies $E_T$, or $\Delta H_T$, of these two pseudojets is minimized. The scalar sum of the transverse energies is defined as
    \begin{equation}
      H_T = \sum^{n_{\mathrm{jet}}}_{i =1} E_T^{j_i},
      \label{eqn:ht_alphat}
    \end{equation}
    with $n_{\mathrm{jet}}$ the number of jets with an $E_T$ above a certain threshold. The definition of $\alpha_T$ then becomes
    \begin{equation}
      \alpha_T = \frac{1}{2} \times \frac{H_T - \Delta H_T}{\sqrt{H_T^2 - \slashed{H}_T^2}} =  \frac{1}{2} \times \frac{1 - (\Delta H_T/H_T)}{\sqrt{1 - (\slashed{H}_T/H_T)^2}}, 
      \label{eqn:alphaT_pseudojet}
    \end{equation}
    where $\slashed{H}_T$ is the is the magnitude of $\vec{\slashed{H}}_T = -\sum_j \vec{p}_{T,\hspace{1mm}j}$. 

    Our implementation of the analysis is based on the signal region of 2--3 jets without a $b$-tagged jet. We use the same binned ranges in $H_T$ as \cite{Chatrchyan:2013lya}, so that we have bin 0 for the range 275--325 GeV, bin 1 for the range 325--375 GeV, bins 2--6 for ranges of 100 GeV between 375--875 GeV, and an open bin 7 with $H_T >$ 875 GeV. We also have a combination of all bins with $H_T > $ 275 GeV, denoted by bin 8. 

\paragraph{The MHT search \cite{Chatrchyan:2014lfa}}
The MHT search, which, like the $\alpha_T$ based analysis, is designed for jets and missing energy, does not include $b$ quarks; this in contrast to the $\alpha_T$ analysis described above. It is based on two variables, the sum of the transverse momenta $H_T$ for jets $j$ defined as 
\begin{equation}
  H_T = \sum_j p_{T,\hspace{1mm}j},\hspace{1cm} p_{T, j} > 50 \hspace{2mm} \mathrm{GeV}; \hspace{2mm} |\eta_{j}| < 2.5,
  \label{eqn:sumtransversemomenta}
\end{equation}
and the vector sum of the missing transverse momenta $\vec{\slashed{H}}_T$ of jets $j$
\begin{equation}
  \vec{\slashed{H}}_T = -\sum_j \vec{p}_{T,\hspace{1mm}j},\hspace{1cm} p_{T, j} > 30 \hspace{2mm} \mathrm{GeV}; \hspace{2mm} |\eta_{j}| < 5.
  \label{eqn:summissingtransversemomenta}
\end{equation}
As before, $\slashed{H}_T$ is the magnitude of $\vec{\slashed{H}}_T$. 

The MHT search region of 3--5 jets is binned into 0--16 bins in the following ranges of the variables $H_T$ and $\slashed{H}_T$ as follows:
\begin{itemize}
  \item 500--800, 800--1000, and 1000--1250 GeV in $H_T$, and 200--300, 300--450, 450--600, and $>$ 600 GeV for $\slashed{H}_T$ (bins 0--11);
  \item 1250--1500 GeV in $H_T$ with $\slashed{H}_T$ binned into 200--300, 300--450, and $>$ 450 GeV (bins 12--14);
  \item $>$ 1500 GeV in $H_T$ with $\slashed{H}_T$ binned into 200--300 and $>$ 300 GeV (bins 15--16).
\end{itemize}
Bin 17 is the combination of all bins with $H_T$ $>$ 500 GeV and $\slashed{H}_T$ $>$ 200 GeV.

\paragraph{Simulation details}
We simulate events for the simplified SUSY-T2 and UED-T2 models for quark partner masses in the range $m_{\tilde{q}/q^{(1)}}$ from 500 to 1600\,GeV, and 
$m_{\rm LSP/LKP}$  from 100 to $(m_{\tilde{q}/q^{(1)}}-200)$\,GeV. The minimal UED model \cite{Cheng:2002ab, Appelquist:2000nn} from \texttt{Feynrules} \cite{Christensen:2008py,Christensen:2009jx} is modified such that the masses of the KK excitations are taken as free parameters. We use \texttt{MadGraph 5.1.12} \cite{Alwall:2014hca} to generate events at parton level, \texttt{Pythia 6.4} \cite{Sjostrand:2006za} for showering, and subsequently \texttt{Delphes 3.0.11} \cite{deFavereau:2013fsa}, which includes \texttt{FastJet} \cite{fastjet}, for detector simulation. We then analyze the event samples using our own implementation of the $\alpha_T$ and MHT searches briefly described above.

\section{Results}\label{sec:results}
In this section we present the efficiencies for the SUSY-T2 and UED-T2 simplified models and quantify how the differences due to spin affect the mass limits. 
We also discuss the difference between the UED-T2 simplified model with a decoupled KK gluon and more general UED models with a finite KK gluon mass and the corresponding additional production modes due to KK gluon exchange. 

\subsection{SUSY-T2 and UED-T2 simplified model efficiencies}

We first compare the signal acceptance times efficiency $\eaxe$ (simply called ``efficiency" in the following) 
obtained from the SUSY-T2 and UED-T2 simplified models, respectively. Note that in these cases only (s)quark-anti(s)quark production plays a role; (s)quark pair production is absent due to decoupled gluinos and KK gluons. Hence, the only differences between UED-T2 and SUSY-T2 are the difference in spin and one extra diagram for SUSY-T2.

\begin{figure}[h]
\centering
\includegraphics[width=0.85\textwidth, page=2]{plots/JS_T2_vs_dqdqbar_inf.pdf}\\
\includegraphics[width=0.85\textwidth, page=1]{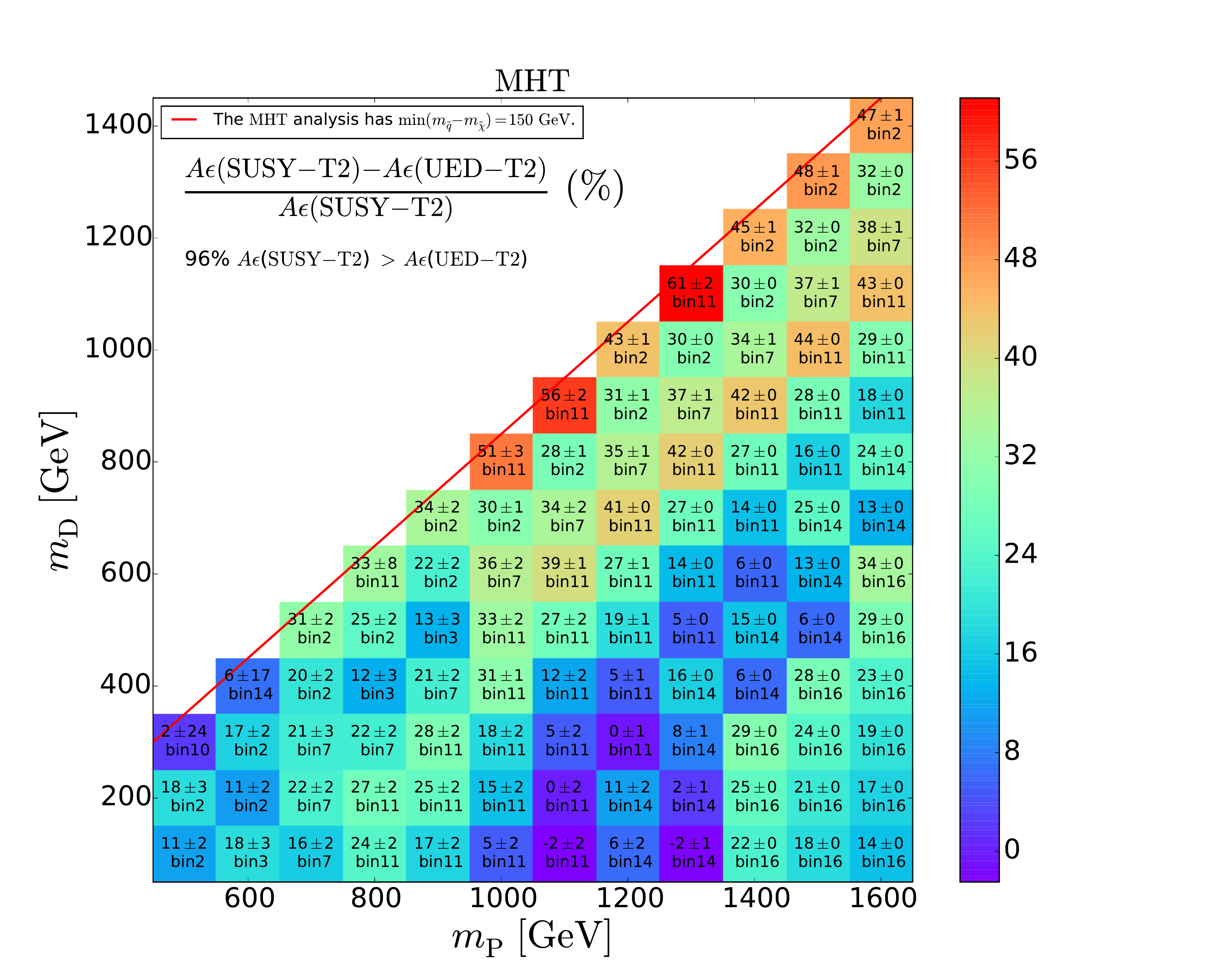}
\caption{Relative differences in efficiencies \axe between the UED-T2 and SUSY-T2 simplified models in the plane of the parent ($q^{(1)}$, $\tilde{q}$) mass $m_{\mathrm{P}}$ and daughter ($\chi_1^0$, $B_1$) mass $m_{\mathrm{D}}$, for the $\alpha_T$ analysis (top) and for the MHT analysis (bottom).}
\label{fig:T2diffaxe}
\end{figure}

We calculate the relative efficiency differences as
\begin{align}
    \Delta_{\eaxe}	&= \frac{\eaxe_{\mathrm{SUSY-T2}}-\eaxe_{\mathrm{UED-T2}}}{\eaxe_{\mathrm{SUSY-T2}}}.
\end{align}
 Figure \ref{fig:T2diffaxe} shows $\Delta_{\eaxe}$ for the $\alpha_T$ analysis (top) and the MHT analysis (bottom). 
 The relative differences are shown for the most sensitive bin, which is the bin yielding the lowest expected upper limit on events for the SUSY-T2 model using a pure background hypothesis.
 The results for the most sensitive bin are also those that have been used to calculate the exclusion contours presented in section~\ref{sec:limits}. 
 The errors shown are the Monte Carlo errors of our simulation.
 The parent particle is $\mbox{P} = \tilde{q},q^{(1)}$ and the daughter particle is $\mbox{D} = \tilde{\chi}, B^{(1)}$ for SUSY-T2 and UED-T2, respectively.
 
 For the $\alpha_T$ search, the \axe of UED-T2 is larger than \axe for SUSY-T2 in most of the investigated parameter space.
 The relative differences between UED-T2 and SUSY-T2 are typically of $\mathcal{O}(25\%)$ with maximal deviations of up to 50\%. 
 Generically,  the relative \axe differences are smaller for small daughter masses $m_{\mathrm{D}}$, and are quite independent from the 
 parent mass $m_{\mathrm{P}}$.   Mass splittings smaller than $m_{\rm P}-m_{\rm D}=150$ GeV are not included in \cite{Chatrchyan:2013lya}.

 For the MHT search, we find that the efficiencies for UED are smaller than those for SUSY-T2, with relative differences of $\mathcal{O}(15-40\%)$ for most points in parameter space, with maximal deviations of up to 60\%.
 Again, mass splittings smaller than $m_{\rm P}-m_{\rm D}=150$ GeV are not included in \cite{Chatrchyan:2014lfa}.

\subsection{Efficiencies for more general UED models}
\label{section:othertops}

As discussed in section \ref{sec:t2setup}, in a general UED model with a finite KK-gluon mass there are additional quark production modes through KK gluon exchange (\ref{eq:ued_prod}), which yield the same jets plus MET signature as the UED-T2 model. 

For the $\alpha_T$ search we find again that the \axe are larger for the UED scenarios than for SUSY-T2, with relative differences typically of $\mathcal{O}(5-30\%)$.  
The largest relative differences in efficiencies from SUSY-T2 are found for $q_{\mathrm{D}}\bar q_{\mathrm{D}}$ production with deviations of up to $\mathcal{O}(50\%)$.
For all channels, deviations were largest for small mass splittings between parent and daughter masses.
The change in relative efficiency differences with a change in the KK gluon mass is of the order of a few percent in these channels. 

In the MHT analysis, the SUSY-T2 simplified model mostly overestimates the efficiencies for UED KK-quark production. The relative differences are largest 
for mixed first- and second-generation and second-generation-only KK quark production, with deviations of up to $\mathcal{O}(60\%)$. The relative differences in efficiencies change from the order of a few percent for most points to $\mathcal{O}( 20\%)$ for a few points when varying the KK gluon mass.

\subsection{Exclusion limits}
\label{sec:limits}
We would like to quantify to what extent the relative differences between the SUSY-T2 and UED signal efficiencies affect the limits on new particle masses. From the 
UED cross section predictions (as obtained with \texttt{MadGraph 5} \cite{Alwall:2014hca}) and the SUSY-T2 and UED efficiencies, respectively, 
we derive  95\% C.L. upper limits on the quark partner mass with \texttt{RooStatsCL95} \cite{Moneta:2010pm}.

\begin{figure}[htp]
\centering
\includegraphics[width=0.45\textwidth]{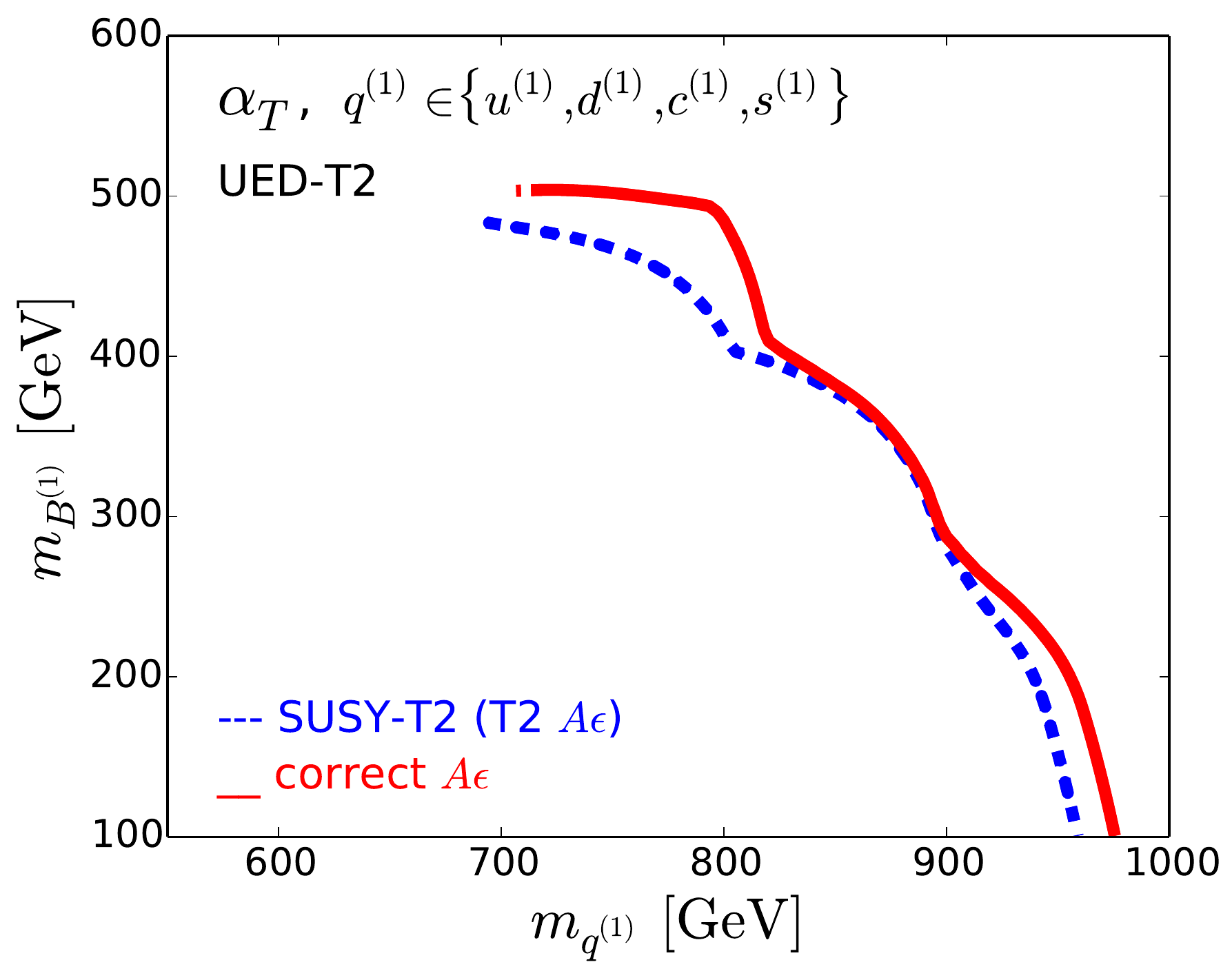}
\includegraphics[width=0.45\textwidth]{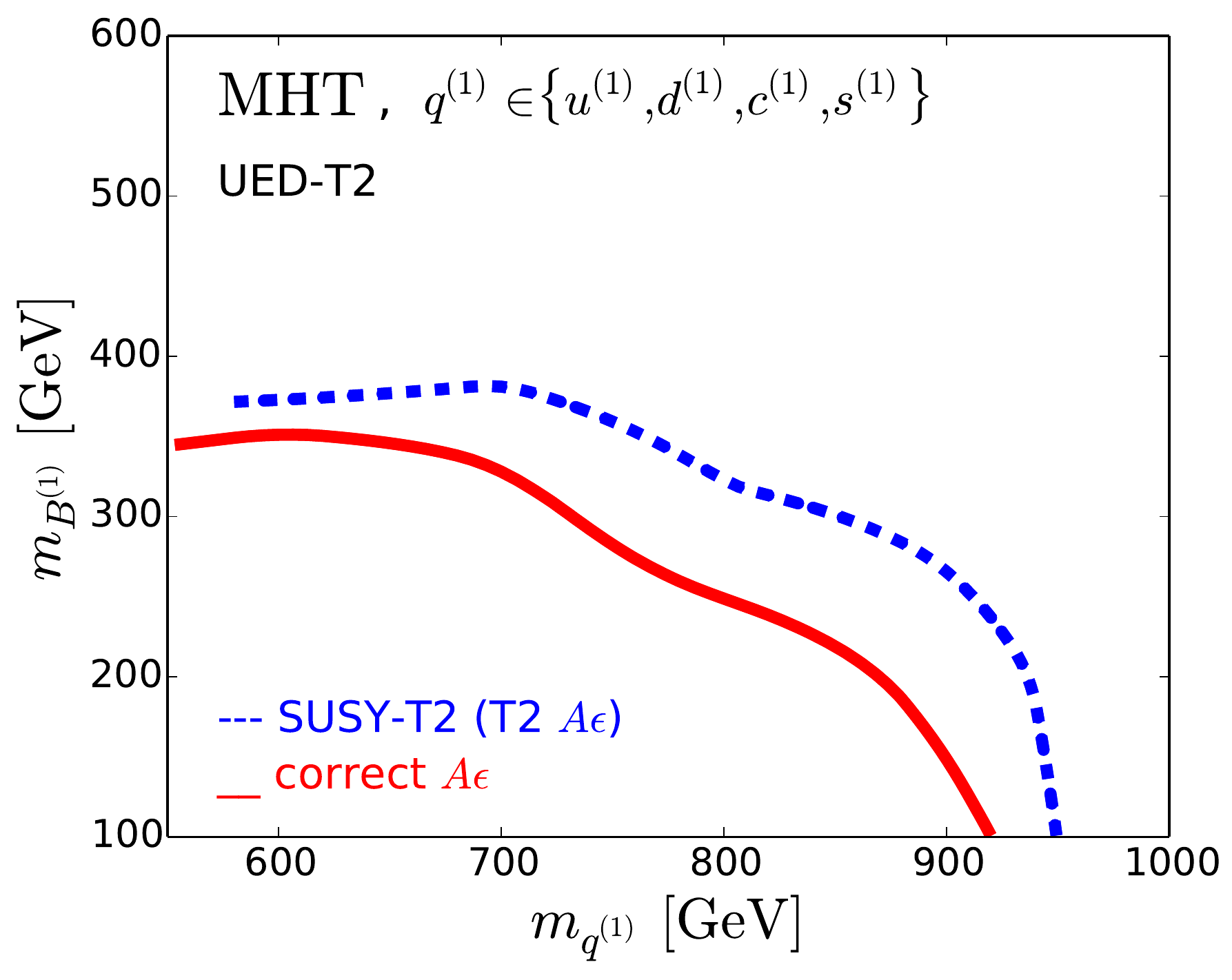}
\caption{95\% C.L. exclusion limits on the KK quark and LKP masses derived from the SUSY-T2 (dashed line) and consistent UED-T2 (solid line) simplified model efficiencies. The $\alpha_T$ and MHT analyses are based on an integrated luminosity of 11.7 fb$^{-1}$ and 19.5 fb$^{-1}$, respectively, collected at the 8 TeV LHC.}
\label{fig:T2difflimits}
\end{figure}

The results of our limit calculation are shown in figure \ref{fig:T2difflimits}.  The red (solid) curves are the consistent limits as obtained for a UED model using the UED-T2 efficiencies, while 
the blue (dashed) curves correspond to inconsistent UED mass limits obtained by using UED cross sections with SUSY-T2 efficiencies.
Since the relative difference $\Delta_{\eaxe}$ is positive for $\alpha_T$, the consistent limit for UED is higher than the limit derived from the efficiencies of SUSY-T2. For the MHT analysis we see the opposite behaviour. The under- (over-) estimation of the limits is smaller (larger) for the $\alpha_T$ (MHT) search, respectively. 
The difference in the quark partner mass limits between the consistent and inconsistent UED interpretations is at most about 70\,GeV and 130\,GeV for the $\alpha_T$ and MHT analyses, respectively, in the mass region probed by current LHC searches. 

In order to investigate limits for a generic UED scenario, we consider a UED KK-quark pair production model with a finite KK-gluon mass and the corresponding efficiency 
\begin{eqnarray}
  \eaxe_{\mathrm{UED}}\times\sigma_{\mathrm{total}} &=&  \eaxe_{q_{\mathrm{D}} q_{\mathrm{D}}}(\sigma_{q_{\mathrm{D}} q_{\mathrm{D}}} + \sigma_{q_{\mathrm{S}} q_{\mathrm{S}}}) +\eaxe_{q_{\mathrm{D}} \bar{q}_{\mathrm{D}}}(\sigma_{q_{\mathrm{D}} \bar{q}_{\mathrm{D}}} +\sigma_{q_{\mathrm{S}} \bar{q}_{\mathrm{S}}}) + \nonumber \\
  & & + \eaxe_{q_{\mathrm{D}} q_{\mathrm{S}}}\sigma_{q_{\mathrm{D}} q_{\mathrm{S}}} + \eaxe_{q_{\mathrm{D}} \bar{q}_{\mathrm{S}}}(\sigma_{q_{\mathrm{D}} \bar{q}_{\mathrm{S}}} + \sigma_{q_{\mathrm{S}} \bar{q}_{\mathrm{D}}}).
  \label{eqn:uedlike}
\end{eqnarray}
Here $\sigma_{\mathrm{total}}$ is the sum of the cross sections of all the KK quark production channels appearing in eq.\,(\ref{eqn:uedlike}).
For second generation KK quarks, one efficiency was calculated for all doublets and singlets, and accordingly 
$\eaxe_{q_{\mathrm{2nd}}q_{\mathrm{both}}}\sigma_{q_{\mathrm{2nd}}q_{\mathrm{both}}}$ was added to the above.
$q_{\mathrm{2nd}}$ are all second generation light KK quarks, and $q_{\mathrm{both}}$ are all first and second generation KK quarks. 

\begin{figure}[htp]
\centering
\includegraphics[width=0.45\textwidth]{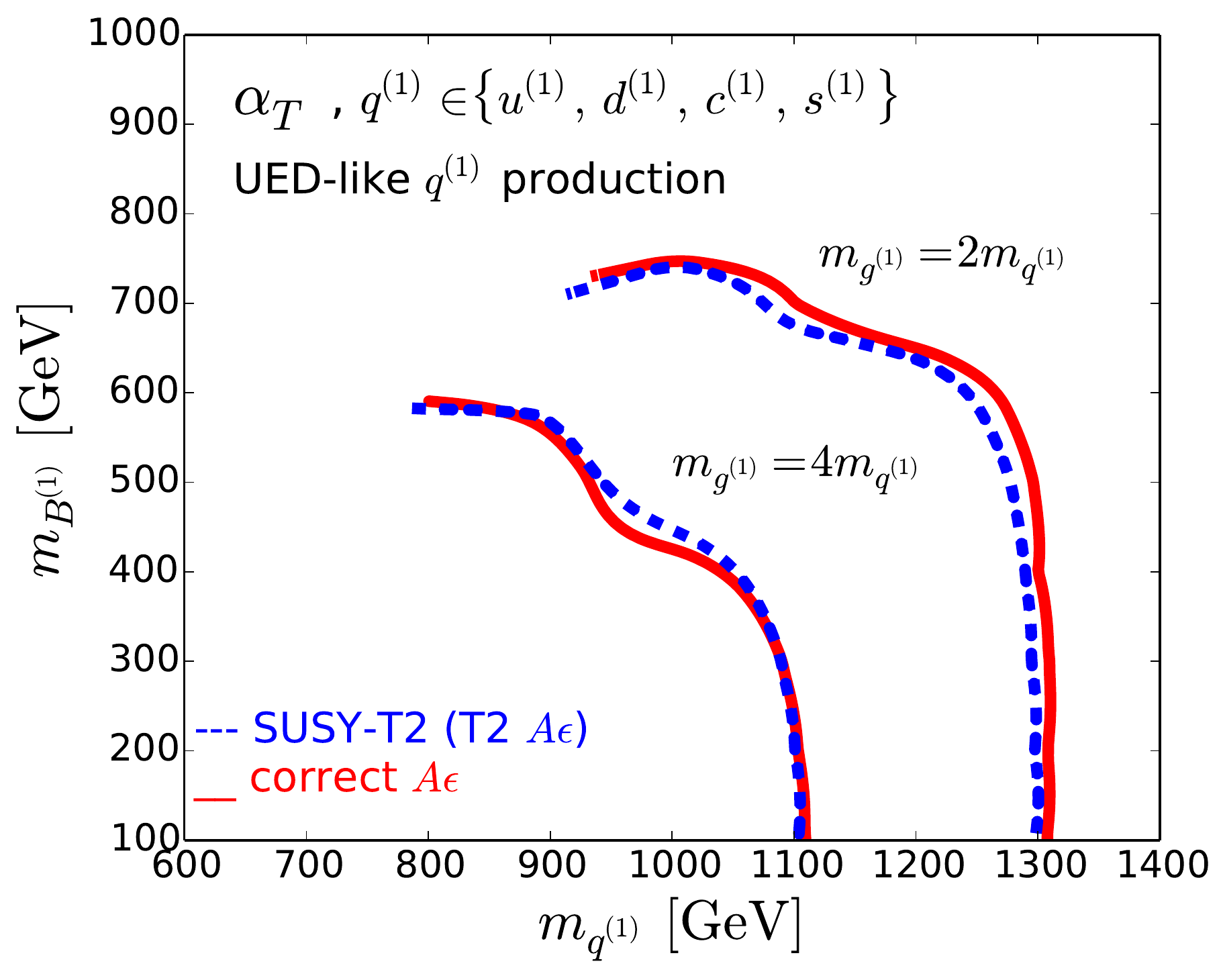}
\includegraphics[width=0.45\textwidth]{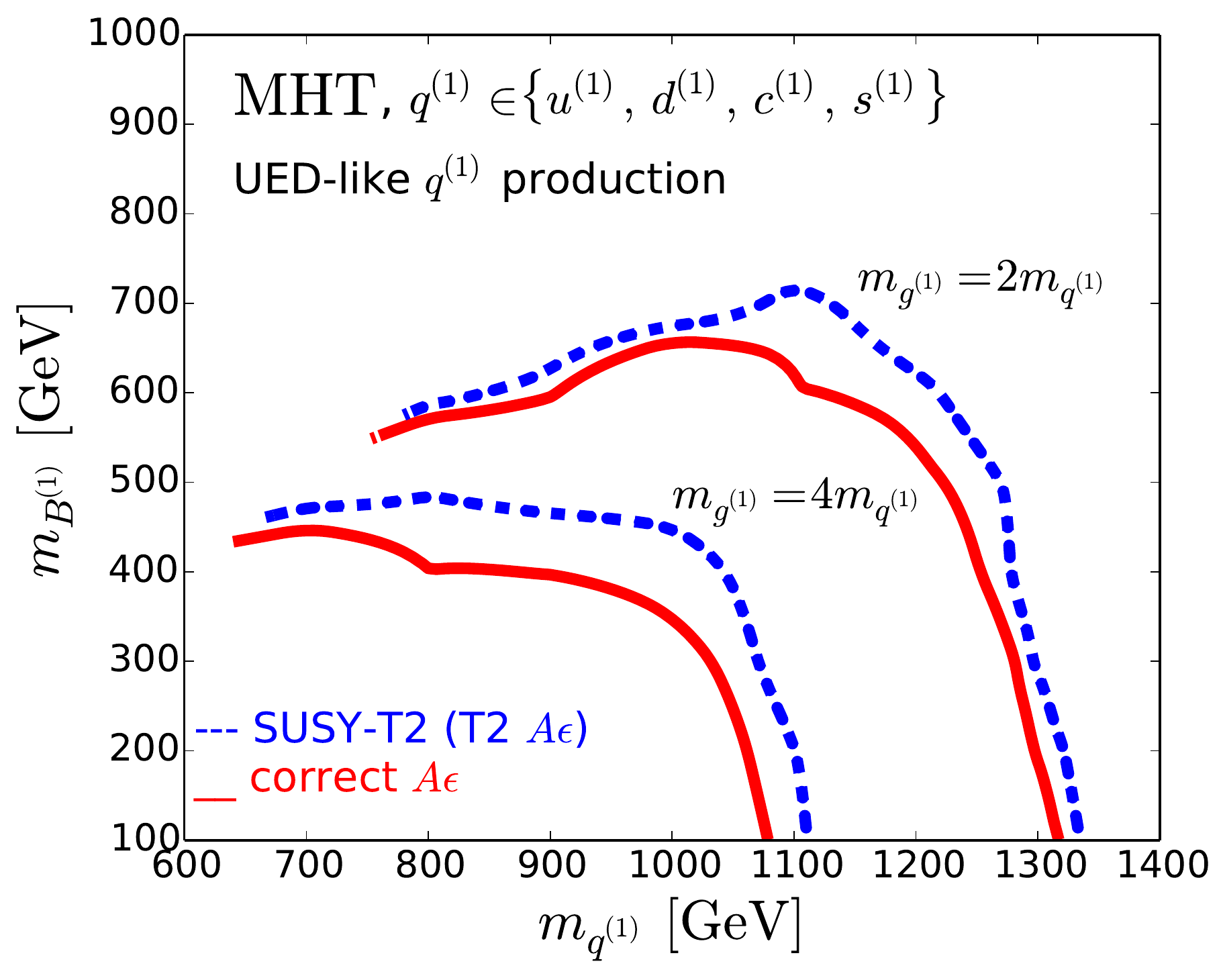}
  \caption{95\% C.L. exclusion limits on the KK quark and LKP masses derived from the SUSY-T2 simplified model (dashed line) and general UED efficiencies (solid line) for a KK gluon of twice and four times the (s)quark mass. 
     The $\alpha_T$ and MHT analyses are based on an integrated luminosity of 11.7 fb$^{-1}$ and 19.5 fb$^{-1}$, respectively, collected at the 8 TeV LHC.}
\label{fig:limitscom2gen}
\end{figure}

Limits for this model are shown in figure \ref{fig:limitscom2gen}. As before, we observe that inconsistently using the SUSY-T2 efficiencies for the UED model yields too weak and too strong limits  in the $\alpha_T$ and MHT analyses, respectively, compared to a consistent interpretation based on the UED efficiencies. However, quantitatively the differences in mass limits derived from the consistent and inconsistent treatment of the efficiencies are insignificant for the $\alpha_T$ analysis, and moderate for MHT, with deviations ranging 
from $\mathcal{O}(20\,{\rm GeV})$ up to $\mathcal{O}(300\,{\rm GeV})$ for certain points in parameter space.

\section{Conclusions}\label{sec:conclusions}
We have investigated how well current experimental searches for new physics based on supersymmetric simplified models can be used to constrain more generic BSM scenarios, like models with same-spin Standard Model partners. 
Specifically, we have considered searches for jets plus missing transverse energy, and interpreted those in terms of simplified models for scalar and fermionic quark-partner pair production with subsequent decays into jets and weakly interacting stable particles.

The spin of the particles affects their kinematic distributions and thus the signal efficiencies and the corresponding exclusion limits. 
We have quantified to what extent the current experimental analyses, which are based on supersymmetric simplified models with scalar quark partners, can be used to constrain same-spin BSM scenarios like models with universal extra dimensions. 
We found sizable differences of up to $\mathcal{O}(60\%)$ in the signal efficiencies of the supersymmetric and universal extra dimension models for certain regions of parameter space. However, the differences in the corresponding mass limits are moderate, with deviations of typically $\mathcal{O}(10\%)$, or 
$\mathcal{O}(100\,\mathrm{GeV})$ in the mass region probed by current LHC searches.
We find that the difference between the true same-spin limits and those obtained by using supersymmetric simplified models is generically smaller for the $\alpha_T$ than for the MHT search. Also, the limits based on supersymmetric simplified models are conservative for the $\alpha_T$ analysis, and slightly too optimistic for the analysis based on MHT. 

We conclude that simplified supersymmetric models provide a reliable and robust tool to analyse the current hadronic jets plus missing energy searches at the LHC. 
The corresponding cross section limits can be interpreted in a wider class of BSM scenarios, including those with same-spin SM partners.

\section*{Acknowledgements}
We would like to thank Jan Heisig, Lennart Oymanns and Wolfgang Waltenberger for useful discussions. We are grateful to Wolfgang and to the Institute of High Energy Physics (HEPHY) for their hospitality during various visits to Vienna. MK is grateful to SLAC and Stanford University for their hospitality during his sabbatical stay. This work was supported by the Deutsche Forschungsgemeinschaft through the research training group ``Particle and Astroparticle Physics in the Light of the LHC" and through the collaborative research centre TTR9 ``Computational Particle Physics", and by the U.S. Department of Energy under contract DE-AC02-76SF00515.


\clearpage
\addcontentsline{toc}{section}{References}
\bibliographystyle{JHEP}
\bibliography{SMSref}

\providecommand{\href}[2]{#2}\begingroup\raggedright\begin{thebibliography}{10}

\bibitem{Alves:2011wf}
{\bf LHC New Physics Working Group} Collaboration, D.~Alves et~al., {\it
  {Simplified Models for LHC New Physics Searches}},
  \href{http://arxiv.org/abs/1105.2838}{{\tt arXiv:1105.2838}}.

\bibitem{Alwall:2008ag}
J.~Alwall, P.~Schuster, and N.~Toro, {\it {Simplified Models for a First
  Characterization of New Physics at the LHC}},  {\em Phys. Rev.} {\bf D79}
  (2009) 075020, [\href{http://arxiv.org/abs/0810.3921}{{\tt
  arXiv:0810.3921}}].

\bibitem{Chatrchyan:2013sza}
{\bf CMS Collaboration} Collaboration, S.~Chatrchyan et~al., {\it
  {Interpretation of Searches for Supersymmetry with Simplified Models}},  {\em
  Phys.Rev.} {\bf D88} (2013), no.~5 052017,
  [\href{http://arxiv.org/abs/1301.2175}{{\tt arXiv:1301.2175}}].

\bibitem{Kraml:2013mwa}
S.~Kraml, S.~Kulkarni, U.~Laa, A.~Lessa, W.~Magerl, et~al., {\it {SModelS: a
  tool for interpreting simplified-model results from the LHC and its
  application to supersymmetry}},  {\em Eur.Phys.J.} {\bf C74} (2014) 2868,
  [\href{http://arxiv.org/abs/1312.4175}{{\tt arXiv:1312.4175}}].

\bibitem{Kraml:2014sna}
S.~Kraml, S.~Kulkarni, U.~Laa, A.~Lessa, V.~Magerl, et~al., {\it {SModelS v1.0:
  a short user guide}},  \href{http://arxiv.org/abs/1412.1745}{{\tt
  arXiv:1412.1745}}.

\bibitem{Papucci:2014rja}
M.~Papucci, K.~Sakurai, A.~Weiler, and L.~Zeune, {\it {Fastlim: a fast LHC
  limit calculator}},  {\em Eur.Phys.J.} {\bf C74} (2014), no.~11 3163,
  [\href{http://arxiv.org/abs/1402.0492}{{\tt arXiv:1402.0492}}].

\bibitem{Buckley:2013jua}
A.~Buckley, {\it {PySLHA: a Pythonic interface to SUSY Les Houches Accord
  data}},  \href{http://arxiv.org/abs/1305.4194}{{\tt arXiv:1305.4194}}.

\bibitem{Beenakker:1996ch}
W.~Beenakker, R.~Hopker, M.~Spira, and P.~Zerwas, {\it {Squark and gluino
  production at hadron colliders}},  {\em Nucl.Phys.} {\bf B492} (1997)
  51--103, [\href{http://arxiv.org/abs/hep-ph/9610490}{{\tt hep-ph/9610490}}].

\bibitem{Sjostrand:2006za}
T.~Sjostrand, S.~Mrenna, and P.~Z. Skands, {\it {PYTHIA 6.4 Physics and
  Manual}},  {\em JHEP} {\bf 0605} (2006) 026,
  [\href{http://arxiv.org/abs/hep-ph/0603175}{{\tt hep-ph/0603175}}].

\bibitem{Chatrchyan:2013lya}
{\bf CMS Collaboration} Collaboration, S.~Chatrchyan et~al., {\it {Search for
  supersymmetry in hadronic final states with missing transverse energy using
  the variables AlphaT and b-quark multiplicity in pp collisions at 8 TeV}},
  {\em Eur.Phys.J.} {\bf C73} (2013) 2568,
  [\href{http://arxiv.org/abs/1303.2985}{{\tt arXiv:1303.2985}}].

\bibitem{Chatrchyan:2014lfa}
{\bf CMS Collaboration} Collaboration, S.~Chatrchyan et~al., {\it {Search for
  new physics in the multijet and missing transverse momentum final state in
  proton-proton collisions at $\sqrt{s}$= 8 TeV}},  {\em JHEP} {\bf 1406}
  (2014) 055, [\href{http://arxiv.org/abs/1402.4770}{{\tt arXiv:1402.4770}}].

\bibitem{CMS-PAS-SUS-13-019}
{\bf CMS Collaboration} Collaboration, {\it {Search for supersymmetry in
  hadronic final states using MT2 with the CMS detector at sqrt(s) = 8 TeV}},
  Tech. Rep. CMS-PAS-SUS-13-019, CERN, Geneva, 2014.

\bibitem{ATLAS-CONF-2013-047}
{\it {Search for squarks and gluinos with the ATLAS detector in final states
  with jets and missing transverse momentum and 20.3 fb$^{-1}$ of $\sqrt{s}=8$
  TeV proton-proton collision data}},  Tech. Rep. ATLAS-CONF-2013-047, CERN,
  Geneva, May, 2013.

\bibitem{Edelhauser:2014ena}
L.~Edelh{\"a}user, J.~Heisig, M.~Kr{\"a}mer, L.~Oymanns, and J.~Sonneveld, {\it
  {Constraining supersymmetry at the LHC with simplified models for squark
  production}},  {\em JHEP} {\bf 1412} (2014) 022,
  [\href{http://arxiv.org/abs/1410.0965}{{\tt arXiv:1410.0965}}].

\bibitem{Appelquist:2000nn}
T.~Appelquist, H.-C. Cheng, and B.~A. Dobrescu, {\it {Bounds on universal extra
  dimensions}},  {\em Phys.Rev.} {\bf D64} (2001) 035002,
  [\href{http://arxiv.org/abs/hep-ph/0012100}{{\tt hep-ph/0012100}}].

\bibitem{Hooper:2007qk}
D.~Hooper and S.~Profumo, {\it {Dark matter and collider phenomenology of
  universal extra dimensions}},  {\em Phys.Rept.} {\bf 453} (2007) 29--115,
  [\href{http://arxiv.org/abs/hep-ph/0701197}{{\tt hep-ph/0701197}}].

\bibitem{Servant:2002aq}
G.~Servant and T.~M. Tait, {\it {Is the lightest Kaluza-Klein particle a viable
  dark matter candidate?}},  {\em Nucl.Phys.} {\bf B650} (2003) 391--419,
  [\href{http://arxiv.org/abs/hep-ph/0206071}{{\tt hep-ph/0206071}}].

\bibitem{Servant:2002hb}
G.~Servant and T.~M. Tait, {\it {Elastic scattering and direct detection of
  Kaluza-Klein dark matter}},  {\em New J.Phys.} {\bf 4} (2002) 99,
  [\href{http://arxiv.org/abs/hep-ph/0209262}{{\tt hep-ph/0209262}}].

\bibitem{Nilles:1983ge}
H.~P. Nilles, {\it {Supersymmetry, Supergravity and Particle Physics}},  {\em
  Phys.Rept.} {\bf 110} (1984) 1--162.

\bibitem{Cacciapaglia:2013wha}
G.~Cacciapaglia, A.~Deandrea, J.~Ellis, J.~Marrouche, and L.~Panizzi, {\it {LHC
  Missing-Transverse-Energy Constraints on Models with Universal Extra
  Dimensions}},  {\em Phys.Rev.} {\bf D87} (2013), no.~7 075006,
  [\href{http://arxiv.org/abs/1302.4750}{{\tt arXiv:1302.4750}}].

\bibitem{Datta:2011vg}
A.~Datta, A.~Datta, and S.~Poddar, {\it {Enriching the exploration of the mUED
  model with event shape variables at the CERN LHC}},  {\em Phys.Lett.} {\bf
  B712} (2012) 219--225, [\href{http://arxiv.org/abs/1111.2912}{{\tt
  arXiv:1111.2912}}].

\bibitem{Alwall:2014hca}
J.~Alwall, R.~Frederix, S.~Frixione, V.~Hirschi, F.~Maltoni, et~al., {\it {The
  automated computation of tree-level and next-to-leading order differential
  cross sections, and their matching to parton shower simulations}},  {\em
  JHEP} {\bf 1407} (2014) 079, [\href{http://arxiv.org/abs/1405.0301}{{\tt
  arXiv:1405.0301}}].

\bibitem{Edelhauser:2013lia}
L.~Edelh{\"a}user, T.~Flacke, and M.~Kr{\"a}mer, {\it {Constraints on models
  with universal extra dimensions from dilepton searches at the LHC}},  {\em
  JHEP} {\bf 1308} (2013) 091, [\href{http://arxiv.org/abs/1302.6076}{{\tt
  arXiv:1302.6076}}].

\bibitem{Christensen:2008py}
N.~D. Christensen and C.~Duhr, {\it {FeynRules - Feynman rules made easy}},
  {\em Comput.Phys.Commun.} {\bf 180} (2009) 1614--1641,
  [\href{http://arxiv.org/abs/0806.4194}{{\tt arXiv:0806.4194}}].

\bibitem{Christensen:2009jx}
N.~D. Christensen, P.~de~Aquino, C.~Degrande, C.~Duhr, B.~Fuks, et~al., {\it {A
  Comprehensive approach to new physics simulations}},  {\em Eur.Phys.J.} {\bf
  C71} (2011) 1541, [\href{http://arxiv.org/abs/0906.2474}{{\tt
  arXiv:0906.2474}}].

\bibitem{Datta:2010us}
A.~Datta, K.~Kong, and K.~T. Matchev, {\it {Minimal Universal Extra Dimensions
  in CalcHEP/CompHEP}},  {\em New J.Phys.} {\bf 12} (2010) 075017,
  [\href{http://arxiv.org/abs/1002.4624}{{\tt arXiv:1002.4624}}].

\bibitem{Cheng:2002ab}
H.-C. Cheng, K.~T. Matchev, and M.~Schmaltz, {\it {Bosonic supersymmetry?
  Getting fooled at the CERN LHC}},  {\em Phys.Rev.} {\bf D66} (2002) 056006,
  [\href{http://arxiv.org/abs/hep-ph/0205314}{{\tt hep-ph/0205314}}].

\bibitem{deFavereau:2013fsa}
{\bf DELPHES 3} Collaboration, J.~de~Favereau et~al., {\it {DELPHES 3, A
  modular framework for fast simulation of a generic collider experiment}},
  {\em JHEP} {\bf 1402} (2014) 057, [\href{http://arxiv.org/abs/1307.6346}{{\tt
  arXiv:1307.6346}}].

\bibitem{fastjet}
M.~Cacciari, G.~P. Salam, and G.~Soyez, {\it {FastJet User Manual}},  {\em
  Eur.Phys.J.} {\bf C72} (2012) 1896,
  [\href{http://arxiv.org/abs/1111.6097}{{\tt arXiv:1111.6097}}].

\bibitem{Moneta:2010pm}
L.~Moneta, K.~Belasco, K.~S. Cranmer, S.~Kreiss, A.~Lazzaro, et~al., {\it {The
  RooStats Project}},  {\em PoS} {\bf ACAT2010} (2010) 057,
  [\href{http://arxiv.org/abs/1009.1003}{{\tt arXiv:1009.1003}}].

\end{thebibliography}\endgroup

\end{document}